\documentclass[12pt]{article} \usepackage{amssymb,latexsym}
\usepackage{amsmath}
 
\columnwidth\textwidth

\begin{document}

\newtheorem{df}{Definition} \newtheorem{thm}{Theorem} \newtheorem{lem}{Lemma}
\newtheorem{rl}{Rule} \newtheorem{assump}{Assumption}
\begin{titlepage}
 
\noindent
 
\begin{center} {\LARGE Incompleteness of measurement apparatuses} \vspace{1cm}

P. H\'{a}j\'{\i}\v{c}ek\\ Institute for Theoretical Physics \\ University of
Bern \\ Sidlerstrasse 5, CH-3012 Bern, Switzerland \\ hajicek@itp.unibe.ch

\vspace{1cm}

June 2016 \\

 \vspace{1cm}
 
PACS number: 03.65.-w, 03.65.Ta, 07.07.Df
 
\vspace*{2cm}
 
\nopagebreak[4]
 
\begin{abstract} A complete apparatus is defined as reacting to every state of
the measured system. Standard quantum mechanics of indistinguishable particles
is shown to imply that apparatuses must be incomplete or else they would be
drowned out by noise. Each quantum observable is then an abstract
representation of many measurement apparatuses, each incomplete in a different
way. Moreover, a measured system must be prepared in a state that is
orthogonal to the states of all particles of the same type in the
environment. This is the main purpose of preparations. A system so prepared is
said to have a ``separation status''. A new, more satisfactory definition of
separations status than the spatial one proposed in previous papers then
results. Conditions are specified under which the particles in the environment
may be ignored as is usually done in the theory of measurement.
\end{abstract}

\end{center}

\end{titlepage}

\section{Introduction} In quantum mechanics, systems of the same type (such as
all electrons or all hydrogen atoms) are absolutely indistinguishable:
\begin{quote} Any [registration] performed on the [composite] quantum system
treats all [indistinguishable] subsystems in the same way, and it is
indifferent to a permutation of the labels that we attribute to the individual
subsystems for computational purposes.
\end{quote} (Peres \cite{peres}, p.\ 126). A difficulty was also mentioned
(p.\ 128 of \cite{peres}) that then arose: measurements on quantum system
${S}$ can be disturbed by remote particles that are of the same type as
${S}$. A solution to the problem based on Cluster Separability Principle was
suggested.

The suggestion was developed into a theory in \cite{hajicek2,hajicek4}. There,
local kind of quantum observables was defined, similar to that introduced in
\cite{wan,wanb} for different purposes and the notion of {\em separation
status} based on the local observables was introduced.

However, the local observables and the corresponding separation status solve
only a part of Peres' problem. The present paper delivers a clearer
description of the problem and proposes a better solution to it. Instead of
introducing new complicated kind of observables, we leave the observables as
they are but allow measuring apparatuses to be incomplete. We can then give a
more general, simpler and more satisfactory definition of separation status.

Next, the fact will be explained that no disturbance by the indistinguishable
particles of the environment is observed in quantum experiments and that
successful theories of these experiments can serenely ignore these
particles. The conditions under which such method works will result from the
explanation.

\section{Born rule} One of the basic assumptions of quantum mechanics is the
Born rule (see, e.g., \cite{peres}, p.\ 54): Let ${\mathsf O}$ be an
observable of quantum system ${S}$ with a discrete non-degenerate spectrum
$\{o_k\}$ and eigenstates $|k\rangle$,
$$
{\mathsf O}|k\rangle = o_k|k\rangle\ .
$$
Let ${S}$ be in a state $|\psi\rangle$ and let the decomposition of the state
into the eigenstates be
$$
|\psi\rangle = \sum_k c_k |k\rangle\ .
$$
Then the probability of registering eigenvalue $o_k$ on the state
$|\psi\rangle$ is $|c_k|^2$.

In fact, this measurement of ${\mathsf O}$ must be done by some apparatus, or
meter, ${\mathcal M}$, say, and the probability is understood as a prediction
of the frequency obtained by repeated registrations by ${\mathcal M}$. We are
going to ask the question for which states $|\psi\rangle$ the Born rule is
true?

We are going to work with more general observables and states. Let ${S}$ be a
quantum system with Hilbert space ${\mathbf H}$. A general state of ${S}$ is a
positive, trace-1 operator ${\mathsf T} : {\mathbf H} \mapsto {\mathbf H}$
(often called ``density matrix''), and we denote the convex set of all state
operators by ${\mathbf T}({\mathbf H})$. A general observable is an $n$-tuple
$\{{\mathsf O}_1,\ldots,{\mathsf O}_n\}$ of commuting self-adjoint operators
${\mathsf O}_k : {\mathbf H} \mapsto {\mathbf H}$, $k = 1,\ldots,n$. Let
$\sigma \subset {\mathbb R}^n$ be the spectrum of $\{{\mathsf
O}_1,\ldots,{\mathsf O}_n\}$, ${\mathcal B}({\mathbb R}^n)$ the set of Borel
subsets of ${\mathbb R}^n$ and let ${\mathsf \Pi}(X)$, $X \in {\mathcal
B}({\mathbb R}^n)$, describe the spectral measure of $\{{\mathsf
O}_1,\ldots,{\mathsf O}_n\}$ (see, e.g., \cite{RS2}). In particular, ${\mathsf
\Pi}(X)$ is an orthogonal projection on ${\mathbf H}$ for each $X \in
{\mathcal B}({\mathbb R}^n)$,
$$
{\mathsf \Pi}(X)^\dagger = {\mathsf \Pi}(X)\ ,\quad {\mathsf \Pi}(X)^2 =
{\mathsf \Pi}(X)\ ,
$$
and satisfies the normalisation condition,
\begin{equation}\label{normaliz} {\mathsf \Pi}({\mathbb R}^n) = {\mathsf 1}\ .
\end{equation} We shall also adopt the notion (see, e.g., \cite{ludwig1}):
every quantum measurement can be split into preparation and registration. Then
the generalized Born rule can be formulated as follows.
\begin{assump}\label{rlborn} The probability ${\mathrm P}$ that a value of
observable $\{{\mathsf O}_1,\ldots,{\mathsf O}_n\}$ within $X \in {\mathcal
B}({\mathbb R}^n)$ will be obtained by a registration on state ${\mathsf T}$
is
\begin{equation}\label{born} {\mathrm P} = tr({\mathsf T}{\mathsf \Pi}(X))\ .
\end{equation}
\end{assump}

In practice, the Born rule means that the relative frequencies of the values
obtained by many registration by the same meter ${\mathcal M}$ on the same
state ${\mathsf T}$ must tend to the probabilities given by the Born rule if
the number or the registration increases.

\begin{df}\label{dfAregO} Given state ${\mathsf T}$ and $X \in {\mathcal
B}({\mathbb R}^n)$, let us denote by $\omega[{\mathcal M},{\mathsf T}](X)$ the
relative frequencies of finding values of observable $\{{\mathsf
O}_1,\ldots,{\mathsf O}_n\}$ within $X$ obtained from many registrations by
meter ${\mathcal M}$ on state ${\mathsf T}$. Let
\begin{equation}\label{born2} \omega[{\mathcal M},{\mathsf T}](X) \mapsto
tr({\mathsf T} {\mathsf \Pi}(X))
\end{equation} for some states ${\mathsf T}$ and all $X \in {\mathcal
B}({\mathbb R}^n)$. Then we say that meter ${\mathcal M}$ measures $\{{\mathsf
O}_1,\ldots,{\mathsf O}_n\}$.
\end{df} Definition \ref{dfAregO} differs from the usual assumption by a
weaker requirement on the states: the frequency agrees with the Born rule on
``some states'' but not necessarily on ``all states''. Indeed, ``all states''
seems to be the understanding by various books, such as \cite{ludwig1,peres}
at least implicitly, and \cite{BLM} quite explicitly (``probability
reproducibility condition'', p.\ 29). Let us consider some examples.
\begin{enumerate}
\item The position $\vec{\mathsf x}$ of particle ${S}$ with Hilbert space
${\mathbf H}$ is a triple of self-adjoint operators and its spectral measure
is described by (in $Q$-representation)
$$
{\mathsf \Pi}(X) = \chi_X(\vec{x})\ ,
$$
where $\chi_X(\vec{x})$ is the characteristic function of $X \in {\mathcal
B}({\mathbb R}^3)$. The spectrum $\sigma_{\vec{x}}$ is ${\mathbb
R}^3$. Usually, the position is registered by some detector with active volume
$D$ (see, e.g., \cite{leo}). If the detector gives a response (clicks) then we
conclude that a particle has been detected inside $D$. If the wave function of
the detected particle is $\psi(\vec{x})$ then the probability that the
particle will be found inside the detector is
$$
{\mathrm P}(D) = \int_{{\mathbb R}^3}d^3x \chi_D(\vec{x})|\psi(\vec{x})|^2\ ,
$$
Hence, ${\mathrm P}(D) = 0$ if $\text{supp}\,\psi(\vec{x}) \cap D =
\emptyset$. The integral on the right-hand side represents the trace
(\ref{born}) with ${\mathsf T} = |\psi\rangle\langle\psi|$.

A better meter ${\mathcal M}$ registering position is composed of several such
sub-detectors, ${\mathcal M}_1,\ldots,{\mathcal M}_n$ with disjoint active
volumes $D_1,\ldots,D_n$ and the frequency $\omega[{\mathcal M},\psi]_k$ that
the particle will be found inside $D_k$ then satisfies
$$
\omega[{\mathcal M},\psi]_k \mapsto \int_{{\mathbb R}^3}d^3x
\chi_{D_k}(\vec{x})|\psi(\vec{x})|^2\ .
$$
The meter does not register the whole spectrum but only the part
$\sigma'_{\vec{x}} \subset \sigma_{\vec{x}}$ defined by
$$
\sigma'_{\vec{x}} = \bigcup_{k=1}^n D_k\ .
$$
Hence, for all states $\psi$ such that
\begin{equation}\label{suppsi} \text{supp}\,\psi(\vec{x}) \subset
\sigma'_{\vec{x}}
\end{equation} the detector satisfies the Born rule for all $X \in {\mathcal
B}({\mathbb R}^3)$ because zero probability for ${S}$ being outside of
$\sigma'_{\vec{x}}$ results from both the Born rule and the registrations by
${\mathcal M}$. However, for the states that do not satisfy Eq.\
(\ref{suppsi}), the meter still gives zero probability for ${S}$ being outside
of $\sigma'_{\vec{x}}$ contradicting the Born rule.
\item Next, consider a meter that can register energy (a proportional counter,
say). It reacts to a particle only if the particle energy is larger than some
threshold. Again, such a meter will not react to some states, here to those
whose wave function in momentum representation has a support that lies under
the threshold. This example shows that the problem need not be caused just by
the geometric arrangement of the experiment as in point 1.
\item The Stern-Gerlach meter (see, e.g., \cite{peres}, p.\ 14) can register
the spin observable only if the particle arriving at it can pass through the
opening between the magnets within a narrow range of directions. Thus, it does
not react to a number of states. This example shows that the problem can arise
even for a meter that registers the whole spectrum.
\end{enumerate} Let us compare this with the well-known cases (see, e.g.,
\cite{BLM}) of meters that do not register the whole spectra. For instance, a
real meter can only discriminate between sufficiently different values of an
observable ${\mathsf O}$ with a continuous spectrum that is, it registers only
some coarse-grained version of the spectrum. Thus, one introduces a finite
partition the space ${\mathbb R}^n$,
$$
{\mathbb R}^n = \bigcup_{l=1}^n X_l\ ,
$$
and defines a new observable with spectrum $\{1,2,\ldots,n\}$ that is easily
constructed from ${\mathsf O}$ (for details, see \cite{BLM}, p.\ 35). Notice
that the idea is to modify the observable so that the correspondence between
observable and meter via the Born rule is improved. The meter then does react
to all states of the system and satisfies the Born rule corresponding to the
corrected observable. Still, the method does not work for the cases above, in
which the Borel sets that are controlled by the apparatus do not cover the
spectrum. Such an apparatus does not react to states corresponding to the part
of the spectrum that is not covered, so that the difficulty with the states
also occurs.

The possibility that a meter may control only a (sometimes rather small)
proper subset of the whole Hilbert space, as the meters of the above examples
do, does not seem to be ever mentioned. This might be due to the belief that,
as in most cases of non-ideal real circumstances, the shortcoming of real
meters is a natural way of practical things which just must be taken properly
into account in each particular instance and that some real meters might be
arbitrarily close to the ideal or, at least, that continuous improvement of
techniques will make meters better. The main aim of the next section is to
show that standard quantum mechanics of indistinguishable particles sets a
theoretical limit to this: a meter that were ideal in this sense would be
unable to register its observable at all.

\section{Incomplete apparatuses} Let us now simplify things by considering
observables described by a single operator ($n = 1$). The foregoing section
motivates the following definition.
\begin{df}\label{dfcomplapp} Let ${S}$ be a quantum system with Hilbert space
${\mathbf H}$ and let observable ${\mathsf O}$ be a s.a.\ operator on
${\mathbf H}$ with spectrum $\sigma$ and spectral measure ${\mathsf \Pi}(X)$,
$X \in {\mathcal B}({\mathbb R})$. Let meter ${\mathcal M}$ register
observable ${\mathsf O}$. We say that ${\mathcal M}$ is {\em complete} if
Equation (\ref{born2}) holds true for all ${\mathsf T} \in {\mathbf
T}({\mathbf H})$ and $X \in {\mathcal B}({\mathbb R})$.

If there is any state ${\mathsf T}$ for which the frequency $\omega[{\mathcal
M},{\mathsf T}]({\mathbb R})$ of registering any value by ${\mathcal M}$ is
zero, meter ${\mathcal M}$ is called {\em incomplete}. Let the subset of
states for which this is the case be denoted by ${\mathbf T}({\mathbf
H})_{{\mathcal M}0}$ and the subset of states for which Equation (\ref{born2})
holds by ${\mathbf T}({\mathbf H})_{\mathcal M}$. The convex set ${\mathbf
T}({\mathbf H})_{\mathcal M}$ is called the {\em domain} of ${\mathcal M}$.
\end{df} Thus, the three examples in the foregoing section describe incomplete
apparatuses. We now prove that a complete meter cannot work.

Let registrations by meter ${\mathcal M}$ be performed on a system ${S}$ with
Hilbert space ${\mathbf H}$. Suppose that meter ${\mathcal M}$ registers
observable ${\mathsf O}$ with spectral measure ${\mathsf \Pi}(X)$, $X \in
{\mathcal B}({\mathbb R})$ and is complete. Then, because of the normalisation
condition (\ref{normaliz}), we must have $tr({\mathsf T} {\mathsf
\Pi}({\mathbb R})) = 1$ for {\em any state} ${\mathsf T}$. This means that
{\em any} registration on ${\mathsf T}$ by ${\mathcal M}$ must give {\em some}
result.

Then, according to the theory of indistinguishable systems, ${\mathcal M}$
must also register some values on any state ${\mathsf T}'$ of any system
${S}'$ of the same type as ${S}$. Clearly, this is a difficulty: the
measurement of observable ${\mathsf O}$ of ${S}$ by ${\mathcal M}$ is
disturbed by the existence of a system of the same type as ${S}$ anywhere else
in the world, even if it is localised arbitrarily far away from ${S}$ because
it cannot be distinguished from ${S}$ by ${\mathcal M}$. In fact, for most
microsystems ${S}$, the world contains a huge number of systems of the same
type so that a horrible noise must disturb any registration by a complete
meter.

To show the problem in more detail, let us consider two distant laboratories,
A and B. Let ${\mathsf O}$ be a non-degenerate discrete observable of ${S}$
with eigenstates $|k\rangle$ and eigenvalues $o_k$. Let state $|k\rangle$ be
prepared in A and $|l\rangle$ in B so that $k \neq l$ and let ${\mathsf O}$ is
registered in laboratory A by complete meter ${\mathcal M}$. Using Fock space
formalism, we have
\begin{equation}\label{FockA} {\mathsf O} = \sum_n o_n{\mathsf
a}^\dagger_n{\mathsf a}_n\ ,
\end{equation} where ${\mathsf a}_k$ is an annihilation operator of state
$|k\rangle$ (see, e.g., \cite{peres}, p.\ 137).  Such an observable perfectly
expresses the fact that the meter cannot distinguish particles of the same
type. The state prepared by the two laboratories is
\begin{equation}\label{Fockstate} {\mathsf a}^\dagger_k{\mathsf
a}^\dagger_l|0\rangle\ .
\end{equation} For the average $\langle {\mathsf O} \rangle$ of (\ref{FockA})
in state (\ref{Fockstate}), the standard theory of measurement gives
$$
\langle {\mathsf O} \rangle = \langle 0|{\mathsf a}_l{\mathsf a}_k\left(\sum_n
a_n{\mathsf a}^\dagger_n{\mathsf a}_n\right){\mathsf a}^\dagger_k{\mathsf
a}^\dagger_l|0\rangle\ .
$$
Using the relation
$$
{\mathsf a}_r{\mathsf a}^\dagger_s = \eta {\mathsf a}^\dagger_s{\mathsf a}_r +
\delta_{rs}\ ,
$$
where $\eta = 1$ for bosons and $\eta = -1$ for fermions, we can bring all
annihilation operators to the right and all creation ones to the left
obtaining
$$
\langle {\mathsf O} \rangle = a_k + a_l\ .
$$
The result is independent of the distance between the laboratories. Thus, the
measurement in A by any complete meter depends on what is done in B.

Let us next suppose that ${\mathcal M}$ is incomplete in such a way that the
state of any system of the same type as ${S}$ that may occur in the
environment of ${S}$ lies in ${\mathbf T}({\mathbf H})_{{\mathcal
M}0}$. Apparently, such an assumption can be checked experimentally by looking
at the level of noise of the meter. Then, if we prepare a copy of system ${S}$
in a state that lies within ${\mathbf T}({\mathbf H})_{\mathcal M}$ the
registration of ${S}$ by ${\mathcal M}$ cannot be disturbed by the systems in
the environment. In fact, this must be the way of how all quantum measurement
are carried out. We can say that objective properties of our environment
require certain kind of incompleteness of registration apparatus ${\mathcal
M}$ in order that ${\mathcal M}$ can work in this environment.

Accordingly, the course of any successful measurement must be as
follows. First, a registration apparatus ${\mathcal M}$ for a system ${S}$
with Hilbert space ${\mathbf H}$ is constructed and checked. In particular,
the level of its noise must be sufficiently low. From the construction of the
meter, we can infer some set ${\mathbf T}_{\mathcal M}$ of states on which the
meter is able make registrations. ${\mathbf T}_{\mathcal M}$ might be smaller
than the whole domain,
$$
{\mathbf T}_{\mathcal M} \subset {\mathbf T}({\mathbf H})_{\mathcal M}
$$
(the domain is often difficult to specify). Second, systems ${S}$ is prepared
in one of such states. The registration by ${\mathcal M}$ will then not be
disturbed and the probabilities of the results can be calculated theoretically
by formula (\ref{born2}).

The states of ${\mathbf T}_{\mathcal M}$ must therefore be in some sense
sufficiently different from the states of all systems of the same type as $S$
that occur in the environment of $S$. Let us try to express this idea
mathematically. This can be done in the simplest way, if we choose a
particular representation so that the wave function of an extremal state
$|\psi\rangle$ will be $\psi(\lambda)$ and the kernel of a state of arbitrary
external object will be
$T(\lambda^{(1)},\ldots,\lambda^{(N)};\lambda^{(1)\prime},\ldots,\lambda^{(N)\prime})$. For
example, $|\psi\rangle$ in $Q$-representation will be $\psi{\vec{x}}$, in
$P$-representation $\tilde{\psi}(\vec{p})$, etc. Then,
$$
\langle \psi|\phi \rangle = \int d\lambda\,\psi^*(\lambda) \phi(\lambda)\ ,
$$
where $\int d\lambda$ is a generalised integral (such as Lebesgue-Stieltjes
one, see \cite{RS1}, p.\ 19, or, in some cases, convolution of distributions,
see \cite{RS1}, p.\ 323).

Then we propose the following definition:
\begin{df}\label{dfss} Let system $S$ with Hilbert space ${\mathbf H}$ be
prepared in state $|\psi\rangle \langle \psi| \in {\mathbf T}({\mathbf
H})$. Let the environment consists of macroscopic objects with quantum models
and well-defined quantum states. Let $O$ be the system associated with such an
object, $O_S$ the subsystem of $O$ containing all subsystems of $O$ that are
indistinguishable from $S$ and let ${\mathsf T}$ be the state of $O_S$. If
\begin{equation}\label{dfsseq} \int d\lambda^{(1)\prime}
T(\lambda^{(1)},\ldots,\lambda^{(N)};
\lambda^{(1)\prime},\ldots,\lambda^{(N)\prime})\psi(\lambda^{(1)\prime}) = 0
\end{equation} holds for any object in the environment, then $|\psi\rangle
\langle \psi|$ is said to have {\em separation status}.
\end{df} The definition can easily be extended to states of $S$ that are not
extremal. Some motivation of the definition is as follows. Suppose that there
is a system in the environment in a state $\phi$ and that $\langle \phi| \psi
\rangle \neq 0$. Then,
$$
\phi = c_1\psi + c_2 \psi^\bot\ ,
$$
with non-zero $c_1$ and $\langle \psi^\bot| \psi \rangle = 0$, and the meter
would react to the $\psi$-part of $\phi$.

The above ideas also have some relevance to the meaning of preparation
processes in quantum measurements. Stating such meaning extends the Minimum
Interpretation, for which preparations and registrations are primitive notions
(see, e.g., \cite{peres}, p.\ 12).
\begin{assump}\label{shprep} Any preparation of a single microsystem must
yield a state having a separation status.
\end{assump}

A separation status of a microsystem is a property that is uniquely determined
by a preparation. Hence, it belongs to objective properties of quantum systems
according to \cite{models,PHJT,hajicek5}. But it is a property that is a
necessary condition for any other objective property because each preparation
must create a separation status. Moreover, only a separation status makes a
quantum system distinguishable from each other system in the environment and
so to a physical object. Thus, a quantum physical object can come into being,
namely in a preparation process, and can expire, viz. if it loses its
separation status.

We can understand the role of incompleteness of meters better if we compare
quantum apparatuses with classical ones. To this aim, we construct a simple
model of an eye. Indeed, an eye is a classical registration device, either by
itself or as a final part of other classical apparatuses.

Our model consists of an optically sensitive surface (retina) that can
register visible light (i.e., with a wave-length between 0.4 and 0.75
$\mu$m). It can distinguish between some small intervals of the visible wave
lengths and between small spots where the retina is hit by light.

The retina covers one side of a chamber that has walls keeping light away
except for a small opening at the side opposite to the retina wall\footnote{An
eye with a small opening instead of a lens occurs in some animals such as
nautilus.}. The radius of the circular opening even if very small is much
larger than the wave length of visible light so that this light waves suffer
only a negligible bending as they pass the opening. The assumed insensitivity
of the retina to smaller wave lengths is an incompleteness that helps to make
the picture sharp.

Another aspect of incompleteness is that only the light that can pass the
opening will be registered. Again, this is important: if the retina were
exposed to all light that can reach it from the neighbourhood, only a smeared,
more or less homogeneous signal would result. Because of the restrictions, a
well-structured colour picture of the world in front of the eye will appear on
the retina.

\section{Observables} There are two ways of how one could react to the
necessary incompleteness of registration apparatuses. First, one can try to
modify the observable that is registered by such a meter so that the results
of the registrations and the probabilities calculated from the Born rule
coincide, similarly as it has been done above for coarse-grained version of
the spectrum. Second, one can leave the observables as they are and accept the
fact that every meter can register its observable only partially. In our
previous work (\cite{hajicek2,hajicek4}), we have tried the first way. It
turned out, however, that the modification that was necessary for an
observable to describe how a real meter worked was messy. Not only the notion
of observable became rather complicated but also only some idealized kinds of
meters could be captured in this way.

The mentioned idealized kind of incomplete meter ${\mathcal M}$ can be
described as follows. Such an ${\mathcal M}$ determines a closed linear
subspace ${\mathbf H}_{\text{ss}}$ of ${\mathbf H}$ so that, instead of
Equation (\ref{born2}), we have
$$
\omega[{\mathcal M},{\mathsf T}](X) \mapsto tr\Biggl(\Bigl({\mathsf
\Pi}_{\text{ss}}{\mathsf T}{\mathsf \Pi}_{\text{ss}}\Bigr){\mathsf
\Pi}(X)\Biggr)
$$
for all ${\mathsf T} \in {\mathbf T}({\mathbf H})$ and $X \in {\mathcal
B}({\mathbb R})$, where ${\mathsf \Pi}_{\text{ss}}$ is the orthogonal
projection onto ${\mathbf H}_{\text{ss}}$. Then,
\begin{equation}\label{specdomain} {\mathbf T}({\mathbf H})_{\mathcal M} =
{\mathbf T}({\mathbf H}_{\text{ss}})
\end{equation} and
$$
{\mathbf T}({\mathbf H})_{{\mathcal M}0} = {\mathbf T}([{\mathsf 1} - {\mathsf
\Pi}_{\text{ss}}]{\mathbf H})
$$
because any element ${\mathsf T}$ of ${\mathbf T}({\mathbf H}_{\text{ss}})$
satisfies
\begin{equation}\label{THA} {\mathsf T} = {\mathsf \Pi}_{\text{ss}}{\mathsf
T}{\mathsf \Pi}_{\text{ss}}\ .
\end{equation}

The construction of the corresponding ``generalized observable'' is
simple. First, we have
$$
tr\Biggl(\Bigl({\mathsf \Pi}_{\text{ss}}{\mathsf T}{\mathsf
\Pi}_{\text{ss}}\Bigr){\mathsf \Pi}(X)\Biggr) = tr\Biggl({\mathsf
T}\Bigl({\mathsf \Pi}_{\text{ss}}{\mathsf \Pi}(X){\mathsf
\Pi}_{\text{ss}}\Bigr)\Biggr)\ .
$$
Next, consider operator ${\mathsf \Pi}_{\text{ss}}{\mathsf \Pi}(X){\mathsf
\Pi}_{\text{ss}}$. It is bounded by ${\mathsf 1}$ and self adjoint because
${\mathsf \Pi}_{\text{ss}}$ and ${\mathsf \Pi}(X)$ are. It is obviously
positive. Thus, it is an {\em effect} (see \cite{BLM,models}). A collection of
effects ${\mathsf E}(X)$, $X \in {\mathcal B}({\mathbb R})$, with certain
properties (including the normalisation condition ${\mathsf E}({\mathbb R}) =
{\mathsf 1}$) is called a {\em positive-operator valued measure} (POVM) (for a
definition see, e.g., \cite{BLM}) and generalizes the notion of spectral
measure. The collection of effects ${\mathsf \Pi}_{\text{ss}}{\mathsf
\Pi}(X){\mathsf \Pi}_{\text{ss}}$ for all $X \in {\mathcal B}({\mathbb R})$ is
not a POVM, however, because we have, instead of the above normalisation
condition,
$$
{\mathsf \Pi}_{\text{ss}}{\mathsf \Pi}({\mathbb R}){\mathsf \Pi}_{\text{ss}} =
{\mathsf \Pi}_{\text{ss}}
$$
Such a quantity could be called ``truncated POVM''. Thus, the notion of
observable had to be changed from a self-adjoint operator to a truncated POVM.

However, the above model of incomplete meter is too simple. For instance, some
of the examples listed in Section 2 cannot be described by it. Indeed,
consider the Stern-Gerlach meter that is arranged in such a way that it can
react to particles moving within a thin tube around the third axis of
coordinates $x_1,x_2,x_3$. The particle that can be registered must thus
arrive at the magnets only within some small subset of the (1,2)-plane, the
third component of its momentum must satisfy
$$
p_3 \in (a_3,b_3)\ ,
$$
which can be large, and
$$
p_1 \in (-c_1,c_1)\ ,\quad p_2 \in (-c_2,c_2)\ ,
$$
where $c_k < \epsilon$ for $k = 1,2$ and for sufficiently small
$\epsilon$. However, these conditions can be satisfied, by any wave packets,
only approximately. Then, the Born rule will also be satisfied only
approximately. Now, a linear superposition of such packets need not be again
such a packet. The above conditions mean that the wave function (in $Q$- or
$P$-representation) of the registered particle must satisfy inequalities of
the form
$$
|\psi(\lambda)|^2 < \epsilon'
$$
for some fixed values of $\lambda$ determined by the arrangement, where
$\lambda$ stands either for $\vec{x}$ or for $\vec{p}$, and $\epsilon'$ is a
small positive number. Suppose that another wave function, $\phi$, also
satisfies the condition. Then it only follows, for all $c$ and $c'$ satisfying
$|c|^2 + |c'|^2 = 1$, that
$$
|c \psi(\lambda) + c' \phi(\lambda)| < 2\epsilon'\ .
$$
Hence, the packets need not form a closed linear subspace of ${\mathbf H}$.

The approach using incomplete meters works even if the domain of an meter does
not satisfy Eq.\ (\ref{specdomain}). In fact, the knowledge of the whole
domain ${\mathbf T}({\mathbf H})_{\mathcal M}$ of an meter is not necessary
for the construction of a model of a registration by it because it is
sufficient to know only those elements of ${\mathbf T}({\mathbf H})_{\mathcal
M}$ that are prepared for the experiment.

These are the reasons why we adopt the second way in the present paper. Then,
the standard notion of observable (a self-adjoint operator) makes the
following sense. We can assume that the union of domains of all possible
meters that can register a given observable covers, in an ideal case, the
whole set of states ${\mathbf T}({\mathbf H})$. For example, proportional
counters can register energy of free particles while meters using scattering
of photons can register energy of its bounded states, etc. Thus, a standard
observable can not be determined by one meter but with all meters that can
measure it. As each of these meters must be incomplete, a number of meters is
needed for one observable.

Everything that has been said in this and the foregoing sections can easily be
extended if the notion of observable is generalized from a self-adjoin
operator to a POVM.

\section{Tensor-product method} We have seen in the foregoing sections that
the disturbance of measurement by environmental particles can be avoided, if
the measuring apparatus is suitably incomplete and the measured system is
prepared in a state with a separation status.

The present section is going to study this in more mathematical detail. In
particular, we shall consider two {\em ways of description} of composite
states. The first way works with the tensor product of the environmental and
the registered system states and the second one with the symmetrized or
anti-symmetrized state of the whole composite system as required by rules of
the theory of indistinguishable systems. On the one hand, the second way of
description is in any case the correct one and we shall have to show that the
two descriptions lead to the same measurable results. On the other hand, the
first way is the only practically feasible one because it does not require the
knowledge of the environment state.

To develop the two descriptions, let us consider system ${S}$ and its
environment ${\mathcal E}$ with the system ${\mathcal E}_{S}$ of all its
subsystems that are indistinguishable from ${S}$. Let $\psi(\lambda)$ be the
wave function of ${S}$, where $\lambda$ is a shorthand for four arguments, for
example three components of position or momentum and one spin variable $m =
-s,\ldots,+s$. As shown in Section 3, $\psi(\lambda)$ can be any
representation of state $|\psi\rangle$. Let ${\mathbf H}$ be the Hilbert space
of ${S}$ and let us assume that ${\mathcal E}_{S}$ consists of $N$ subsystems
so that the Hilbert space of ${\mathcal E}_{S}$ is ${\mathbf H}^N_\tau$. Here
$\tau$ takes value $-1$ for fermions and $+1$ for bosons and as index in the
expression ${\mathbf H}^N_\tau$ it denotes the antisymmetrization (for $\tau =
-1$) or symmetrization (for $\tau = +1$) of the tensor product of $N$ copies
of ${\mathbf H}$. If we are going to speak about both cases, we use the
expression ``$\tau$-symmetrization''. A wave function of ${\mathcal E}_{S}$
has the form
$$
\Psi(\lambda^{(1)},\ldots,\lambda^{(N)}) \in {\mathbf H}^N_\tau
$$
and is $\tau$-symmetric in all arguments
$\lambda^{(1)},\ldots,\lambda^{(N)}$. Then the wave functions of the two
descriptions are
\begin{equation}\label{firstway} \Psi(\lambda^{(1)}, \ldots, \lambda^{(N)})
\psi(\lambda^{(N+1)})
\end{equation} and
\begin{equation}\label{secndway} N_{\text{exch}}{\mathsf
\Pi}^{N+1}_\tau\Bigl(\Psi(\lambda^{(1)}, \ldots, \lambda^{(N)})
\psi(\lambda^{(N+1)})\Bigr)\ ,
\end{equation} where ${\mathsf \Pi}^{N+1}_\tau : {\mathbf H}^N_\tau \otimes
{\mathbf H} \mapsto {\mathbf H}^{N+1}_\tau$ is the orthogonal projection onto
the $\tau$-symmetrized subspace and $N_{\text{exch}}$ is a suitable
normalization factor.

To explain the projection ${\mathsf \Pi}^{N+1}_\tau$ and the normalization
factor $N_{\text{exch}}$ (for details and general proofs, see e.g.,
\cite{models}), we choose $N = 2$ and consider ${\mathsf \Pi}^2_\tau :
{\mathbf H} \otimes {\mathbf H} \mapsto {\mathbf H}^2_\tau$. Let
$\Psi(\lambda^{(1)},\lambda^{(2)}) \in {\mathbf H}^2$, then
\begin{eqnarray*} {\mathsf \Pi}^2_- \Psi(\lambda^{(1)},\lambda^{(2)}) &=&
\frac{1}{2}[\Psi(\lambda^{(1)},\lambda^{(2)}) -
\Psi(\lambda^{(2)},\lambda^{(1)})]\ , \\ {\mathsf \Pi}^2_+
\Psi(\lambda^{(1)},\lambda^{(2)}) &=&
\frac{1}{2}[\Psi(\lambda^{(1)},\lambda^{(2)}) +
\Psi(\lambda^{(2)},\lambda^{(1)})]\ .
\end{eqnarray*} Orthogonal projections do not preserve the
normalization. Hence, the projection must be followed by a normalization
factor, which we will denote by $N_{\text{exch}}$ standing before the
projection symbol. Of course, $N_{\text{exch}}$ depends on the projection and
the wave function being projected, but we just write $N_{\text{exch}}$ instead
of $N_{\text{exch}}({\mathsf \Pi}^2_\tau, \Psi)$ to keep equations short.

As $\Psi$ is already $\tau$-symmetric and normalised, the expression
(\ref{secndway}) can be rewritten as follows:
\begin{multline}\label{secway} N_{\text{exch}}{\mathsf
\Pi}^{N+1}_\tau\Bigl(\Psi(\lambda^{(1)}, \ldots, \lambda^{(N)})
\psi(\lambda^{(N+1)})\Bigr) \\ = N'\sum_{K=1}^{N+1}(\tau)^{N+1-K}
\Psi(\lambda^{(K+1)},\ldots,\lambda^{(N+1)},\lambda^{(1)},\ldots,\lambda^{(K-1)})\psi(\lambda^{(K)})\
.
\end{multline} This relation will simplify some subsequent calculations.

Equation (\ref{secndway}) shows that we can recover the second description
from the first one, but if the two descriptions are to be equivalent in any
sense, one had to recover the first one from the second, too. For this aim,
the separation status is necessary. Let state $\psi(\lambda)$ be prepared with
separation status and let ${\mathsf \Pi}_\psi = |\psi\rangle \langle
\psi|$. Eq.\ (\ref{dfsseq}) then implies
$$
N' = \frac{1}{\sqrt{N+1}}\ .
$$

Now, we make use the fact that the operators on ${\mathbf H}$ can act on
different wave functions (elements of ${\mathbf H}$) in a product and that
this action can be specified by the argument of the function. For example, if
we have product $\psi_1(\lambda^{(1)})\psi_2(\lambda^{(2)})$ and operator
${\mathsf O} : {\mathbf H} \mapsto {\mathbf H}$, operator ${\mathsf O}^{(1)} :
{\mathbf H} \otimes {\mathbf H} \mapsto {\mathbf H} \otimes {\mathbf H}$ is
defined by
$$
{\mathsf O}^{(1)}\Bigr[\psi_1(\lambda^{(1)})\psi_2(\lambda^{(2)})\Bigl] =
({\mathsf O}\psi_1)(\lambda^{(1)})\psi_2(\lambda^{(2)})
$$
while ${\mathsf O}^{(2)}$ by
$$
{\mathsf O}^{(2)}\Bigr[\psi_1(\lambda^{(1)})\psi_2(\lambda^{(2)})\Bigl] =
\psi_1(\lambda^{(1)})({\mathsf O}\psi_2)(\lambda^{(2)})\ .
$$
From the definition of separation status we then obtain
$$
{\mathsf \Pi}^{(k)}_\psi\Psi((\lambda^{(1)}),\ldots,(\lambda^{(N)})) = 0
$$
for any $k = 1,\ldots,N$. With this notation, we can achieve our aim:
obviously,
\begin{multline}\label{piA} {\mathsf
\Pi}^{(N+1)}_\psi\Biggl(N_{\text{exch}}{\mathsf
\Pi}^{N+1}_\tau\Bigl(\Psi(\lambda^{(1)}, \ldots, \lambda^{(N)}) \otimes
\psi(\lambda^{(N+1)})\Bigr)\Biggr) \\ = \nu_\psi\Psi(\lambda^{(1)}, \ldots,
\lambda^{(N)}) \otimes \psi(\lambda^{(N+1)})\ ,
\end{multline} where $\nu_\psi$ is again a suitable normalization
factor. Observe that this operation is naturally described by the formalism of
$\tau$-symmetrized wave functions rather than by the Fock-space formalism. The
exchange symmetry is not violated because we can use ${\mathsf
\Pi}^{(K)}_\psi$ for any fixed $K = 1,\dots, N$ instead of ${\mathsf
\Pi}^{(N+1)}_\psi$ and the result will again be the above tensor product with
renamed arguments.

The next point is to give an account of registration by an incomplete
meter. We construct two observables that is registered by the meter, each for
one of the two description ways, and show that the two ways of descriptions
lead to the same results. We work with a simple model to show the essential
points; the general situation can be dealt with in an analogous way.

Let meter ${\mathcal M}$ register observable ${\mathsf O} : {\mathbf H}
\mapsto {\mathbf H}$ that is additive, discrete and non-degener\-ate. Let its
eigenvalues be $o_k$ and eigenvectors be $\psi_k$, $k \in {\mathbb N}$. Let
${\mathcal M}$ be incomplete in the way that it reacts only to $\psi_k$ if $k
= 1,\ldots,K$ for some $K \in {\mathbb N}$. Hence, the subspace ${\mathbf
H}_{\text{ss}}$ is spanned by vectors $\psi_k$, $k = 1,\ldots,K$, and the
projection onto it is
$$
{\mathsf \Pi}_{\text{ss}} = \sum_{k=1}^K{\mathsf \Pi}_k\ ,
$$
where
$$
{\mathsf \Pi}_k = |\psi_k\rangle \langle \psi_k|\ .
$$
The action of the meter can now be described as follows. Let us prepare state
$\psi$ with a separation status. Hence, $\psi \in {\mathbf H}_{\text{ss}}$ and
its decomposition into the eigenstates of ${\mathsf O}$ is
$$
\psi = \sum_{k=1}^K c_k\psi_k
$$
with $\sum_{k=1}^K |c_k|^2 = 1$. Then the probability ${\mathrm P}_k$ of
registering $o_k$ on $\psi$ is
$$
{\mathrm P}_k = \langle \psi|{\mathsf \Pi}_k |\psi\rangle|^2\ .
$$
In this way,
$$
{\mathrm P}_k = |c_k|^2
$$
for $k \leq K$ and ${\mathrm P}_k = 0$ for $k > K$.

Let us start with the first way, Equation (\ref{firstway}). We define the
corresponding observable by restricting the action of ${\mathsf O}$ or
${\mathsf \Pi}_k$ to the second factor:
\begin{multline}\label{obsfirstway} ({\mathsf 1} \otimes {\mathsf
\Pi}_k)\Bigl[\Psi(\lambda^{(1)}, \ldots, \lambda^{(N)}) \otimes
\psi(\lambda^{(N+1)})\Bigr] \equiv {\mathsf
\Pi}_k^{(N+1)}\Bigl[\Psi(\lambda^{(1)}, \ldots, \lambda^{(N)})
\psi(\lambda^{(N+1)})\Bigr] \\ = c_k\Psi(\lambda^{(1)}, \ldots, \lambda^{(N)})
\psi_k(\lambda^{(N+1)})
\end{multline} for $k \leq K$. Eq.\ (\ref{obsfirstway}) specifies the Born
rule of the observable. Now, coming to the second way of description, Equation
(\ref{secndway}), we use the fact that the observable is additive. For
example, it acts on product $\phi_1(\lambda^{(1)})\phi_2(\lambda^{(2)})$ as
follows
$$
({\mathsf O}^{(1)} + {\mathsf
O}^{(2)})\Bigl(\phi_1(\lambda^{(1)})\phi_2(\lambda^{(2)})\Bigr)\ .
$$
Then, to define the observable registered by ${\mathcal M}$, we need the
action of its projection ${\mathsf \Pi}'_k$ for eigenvalue $o_k$, $k \in
{\mathbb N}$. Let us choose:
\begin{equation}\label{piprimek} {\mathsf \Pi}'_k = \sum_{l=1}^{N+1}({\mathsf
\Pi}_k{\mathsf \Pi}_{\text{ss}})^{(l)}
\end{equation} Observe that operators ${\mathsf \Pi}_k$ and ${\mathsf
\Pi}_{\text{ss}}$ commute. Then, using Eq.\ (\ref{secway}), we obtain
\begin{multline*} \sum_{l=1}^{N+1}({\mathsf \Pi}_k{\mathsf
\Pi}_{\text{ss}})^{(l)}\Bigl[N_{\text{exch}}{\mathsf
\Pi}^{N+1}_\tau\Bigl(\Psi(\lambda^{(1)}, \ldots, \lambda^{(N)})
\psi(\lambda^{(N+1)})\Bigr)\Bigr]= \sum_{l=1}^{N+1}({\mathsf \Pi}_k{\mathsf
\Pi}_{\text{ss}})^{(l)} \\
\left[\frac{1}{\sqrt{N+1}}\sum_{K=1}^{N+1}(\tau)^{N+1-K}
\Psi(\lambda^{(K+1)},\ldots,\lambda^{(N+1)},\lambda^{(1)},\ldots,\lambda^{(K-1)})\psi(\lambda^{(K)})\right]
\\ = \frac{1}{\sqrt{N+1}}\sum_{K=1}^{N+1}(\tau)^{N+1-K}
\Psi(\lambda^{(K+1)},\ldots,\lambda^{(N+1)},\lambda^{(1)},\ldots,\lambda^{(K-1)})c_k\psi_k(\lambda^{(K)})
\\ = c_kN_{\text{exch}}{\mathsf \Pi}^{N+1}_\tau\Bigl(\Psi(\lambda^{(1)},
\ldots, \lambda^{(N)}) \psi_k(\lambda^{(N+1)})\Bigr)
\end{multline*} because operator $({\mathsf \Pi}_k{\mathsf
\Pi}_{\text{ss}})^{(l)}$ annihilates the state to the right if the argument
$\lambda^{(l)}$ is in function $\Psi$ and gives $c_k\psi_k(\lambda^{(l)})$ if
the argument is in $\psi$.  Thus, the Born rules for the observables of the
two ways of description coincide.

In general, the operator (\ref{piprimek}) is not a projection because the
product
$$
({\mathsf \Pi}_k{\mathsf \Pi}_{\text{ss}})^{(r)}({\mathsf \Pi}_k{\mathsf
\Pi}_{\text{ss}})^{(s)}
$$
does not in general vanish for $r \neq s$ and then $({\mathsf \Pi}'_k)^2 \neq
{\mathsf \Pi}'_k$. However, on the subspace of ${\mathsf
\Pi}^{N+1}_\tau({\mathbf H}^N_\tau \otimes {\mathbf H})$ with which we are
working, the product is non-zero only if $r = s$, so that it is a projection
under these conditions.

The last question is whether the dynamical evolutions for the two ways of
description are compatible. First, we define the corresponding
Hamiltonians. Let ${\mathsf H} : {\mathbf H}^N_\tau \otimes {\mathbf H}
\mapsto {\mathbf H}^N_\tau \otimes {\mathbf H}$ be a Hamiltonian for the first
way of description and let us assume that
$$
{\mathsf H}{\mathsf \Pi}^{N+1}_\tau = {\mathsf \Pi}^{N+1}_\tau{\mathsf H}\ .
$$
Such a Hamiltonian leaves the subspace ${\mathsf \Pi}^{N+1}_\tau ({\mathbf
H}^N_\tau \otimes {\mathbf H})$ invariant and can also be viewed as a
Hamiltonian for the second way of description. Then, the two
Schr\"{o}dinger equations that we are going to compare are:
\begin{equation}\label{schrodfirstway} {\mathbf H}[\Psi(\lambda^{(1)}, \ldots,
\lambda^{(N)}) \psi(\lambda^{(N+1)})] = i\hbar\frac{\partial}{\partial
t}[\Psi(\lambda^{(1)}, \ldots, \lambda^{(N)}) \psi(\lambda^{(N+1)}]
\end{equation} for the first way of description and
\begin{equation}\label{schrodsecndway} {\mathbf H}{\mathsf
\Pi}^{N+1}_\tau[\Psi(\lambda^{(1)}, \ldots, \lambda^{(N)})
\psi(\lambda^{(N+1)}] = i\hbar\frac{\partial}{\partial t}{\mathsf
\Pi}^{N+1}_\tau[\Psi(\lambda^{(1)}, \ldots, \lambda^{(N)})
\psi(\lambda^{(N+1)}]
\end{equation} for the second way.

We are able to prove the compatibility only if the evolution preserves the
separation status. Mathematically, this means that the Hamiltonian must
commute with the projections defining the status:
\begin{equation}\label{hamss} {\mathsf H}{\mathsf \Pi}^{(k)}_{\text{ss}} =
{\mathsf \Pi}^{(k)}_{\text{ss}}{\mathsf H}
\end{equation} for all $k = 1,\ldots,N+1$. Then, the projections are conserved
and their eigenspaces are stationary. In the case under study, this implies
that the time derivative commutes with the projections, too:
\begin{equation}\label{timess} \frac{\partial}{\partial t}{\mathsf
\Pi}^{(k)}_{\text{ss}} = {\mathsf
\Pi}^{(k)}_{\text{ss}}\frac{\partial}{\partial t}
\end{equation} for all $k = 1,\ldots,N+1$.

Now, the proof of the compatibility is very simple: applying projection
${\mathsf \Pi}^{(N+1)}_{\text{ss}}$ to both sides of equation
(\ref{schrodsecndway}) and using equations (\ref{piA}), (\ref{hamss}),
(\ref{timess}), we obtain equation (\ref{schrodfirstway}).

For processes, in which e.g.\ a measured system loses its separation status,
the two evolutions are not compatible and the second way equation must be
used. Such processes occur during registration of which many examples have
been given in \cite{hajicek4}. We shall adapt the examples to the new
definition of separation status in another paper.

\section{Conclusion an outlook} From quantum mechanical theory of
indistinguishable particles, a strong disturbance of measurement would follow
for measurements by meters that were complete. Suitably incomplete apparatuses
can measure, if the measured systems are prepared in states with a separation
status. The incomplete apparatus gives probability zero to all values that
could be measured on states of the environment. Then, the environmental
particles that are indistinguishable from the measured system can be ignored,
in both the practice of measurements and in their theoretical treatment, as it
is usually done.

A new definition of separation status is proposed that is different from that
of \cite{hajicek2,hajicek4}. In such a way, some problems of the old
definition are removed and the new notion of separation status is even more
general and simpler to use than the old one. Each preparation must create a
separation status.

The environmental particles that are indistinguishable from the measured
system cannot, however, be ignored in the Schr\"{o}dinger equation if the
evolution does not preserve the separation status. The processes of
separation-status change will find application in our theory of state
reduction \cite{haj2} similar to that described in \cite{hajicek2}.

The attempts to achieve a close relation between a quantum observable and its
measurement apparatus is abandoned. The observables are defined as in the
standard quantum theory. In this way, the simplicity and elegance of the
standard theory of quantum observables is preserved. However, each observable
of a system represents a whole class of apparatuses, each registering only a
part of it. The class satisfies the condition that the union of their domains
contains all states of the system.

The new theory is logically consistent with the rest of quantum mechanics,
agrees with the results of real measurements and our understanding of
measurement apparatuses as well as that of preparation processes is improved.

\subsection*{Acknowledgements} The author is indebted to J\"{u}rg
Fr\"{o}hlich, Stefan J\'{a}no\v{s}, Petr Jizba, Franck Lalo\"{e},
Ji\v{r}\'{\i} Tolar and Howard M. Wiseman for useful discussions.


\begin{thebibliography}{99}
\bibitem{peres}Peres, A., {\em Quantum Theory: Concepts and Methods}; Kluwer:
Dordrecht, 1995.
\bibitem{hajicek2}P. H\'{a}j\'{\i}\v{c}ek, Found.\ Phys., {\bf 41} (2011) 640.
\bibitem{hajicek4}P. H\'{a}j\'{\i}\v{c}ek, Found.\ Phys., {\bf 42} (2012) 555.
\bibitem{wan}Wan, K. K.; McLean, R. G. D., J. Phys.\ A: Math.\ Gen. {\bf 17}
(1984) 837.
\bibitem{wanb}Wan, K. K., {\em From Micro to Macro Quantum Systems. A Unified
Formalism with Superselection rules and its Applications}; Imperial College
Press: London, 2006.
\bibitem{RS2}Reed, M. and Simon, B, {\em Methods of Mathematical Physics II:
Fourier Analysis, Self-Adjointness} Academic Press: San Diego, 1975.
\bibitem{ludwig1}Ludwig, G., {\em Foundations of Quantum Mechanics I};
Springer: New York, 1983. {\em Foundations of Quantum Mechanics II}; Springer:
New York, 1985.
\bibitem{BLM}P. Busch, P. J. Lahti and P. Mittelstaedt, {\it The Quantum
Theory of Measurement}; Springer: Heidelberg, 1996.
\bibitem{leo}Leo, W. R., {\em Techniques for Nuclear and Particle Physics
Experiments}; Springer: Berlin, 1987.
\bibitem{RS1}Reed, M. and Simon, B, {\em Methods of Mathematical Physics I:
Functional Analysis} Academic Press: San Diego, 1980.
\bibitem{models}H\'{a}j\'{\i}\v{c}ek, P., Entropy {\bf 15} (2013) 789.
\bibitem{PHJT}H\'{a}j\'{\i}\v{c}ek, P.; Tolar, J., Found.\ Phys. {\bf 39}
(2009) 411.
\bibitem{hajicek5}P. H\'{a}j\'{\i}\v{c}ek, J. Phys.: Conf.\ Ser.\ {\bf 442}
(2013) 012043.
\bibitem{haj2}P. H\'{a}j\'{\i}\v{c}ek, {\em The phenomenon of state
reduction}, preprint, arxiv:1311.3408
\end{thebibliography}
\end{document}